\documentstyle[preprint,prl,aps]{revtex}
\begin{document}
\title{Ballistic annihilation kinetics for a multi-velocity one-dimensional
ideal gas.}
\author{Michel Droz\cite{ack_uno} and Pierre-Antoine Rey}
\address{D\'epartement de Physique Th\'eorique, Universit\'e de Gen\`eve,
CH 1211 Gen\`eve 4, Switzerland.}
\author{Laurent Frachebourg\cite{ack_due}}
\address{Department of Physics, Boston University, Boston, MA 02215, USA}
\author{Jaros\l aw Piasecki\cite{ack_tre}}
\address{Institute of Theoretical Physics, Warsaw University, Ho\.za 69,
P-00 681 Warsaw, Poland}

\date{\today}
\maketitle
\begin{abstract}
Ballistic annihilation kinetics for a multi-velocity one-dimensional ideal gas
is studied in the framework of an exact analytic approach.  For an initial
symmetric three-velocity distribution, the problem can be solved exactly and it
is shown that different regimes exist depending on the initial fraction of
particles at rest.  Extension to the case of a n-velocity distribution is
discussed.

\vspace {0.3truecm}
{PACS numbers: 82-20.-w, 05.20Dd, 05.40+j}
\end{abstract}
\newpage
\section{Introduction.}

Ballistically-controlled reactions provide simple examples of
non-equilibrium systems with complex kinetics.  In the one-dimensional case,
one considers point particles which move freely, with a
given velocity.  When two particles collide they instantaneously annihilate
each other and disappear from the system. The system with only two possible
velocities $+c$ or $-c$ has been studied in a pioneering work by Elskens and
Frisch~\cite{EF}.  Using combinatorial analysis, they showed that the density
of
particles was decreasing according to a power law ($t^{-1/2}$) in the case of a
symmetric initial velocity distribution.  Later, Krug and Spohn~\cite{KS}
obtained independently similar results.  Recently, Redner and
co-workers~\cite{BRL,R} have studied (essentially numerically) the problem with
more general velocity distributions.  However, no rigorous analytical approach
has been elaborated until one of us (JP) developed a formalism in which the
annihilation dynamics reduced exactly to a single closed equation for the
two-particle conditional probability~\cite{pias}.  The purpose of this work is
to present a method which permits to solve this evolution equation for
discrete velocity distributions.
More precisely, we shall consider the three-velocity case where the initial
velocity distribution $\phi(v;t=0)$ is given by
\begin{equation}
\phi(v;0)=p_+\delta(v-c)+p_0\delta(v)+p_-\delta(v+c)\label{eq:disv}
\end{equation}
with $p_+=p_-$ (symmetric case), and $p_++p_0+p_-=1$.

For the ballistic motion, the collision frequency between two particles is
proportional to their relative velocity.  Then, collisions involving particles
at rest occur less often than the ones involving moving particles.  Indeed, the
collisions between one particle of velocity $+c$ and one of velocity $-c$
(relative velocity $2c$) occur twice more often than those between a particle
at rest and a moving one (relative velocity $c$). As a consequence, one can
expect that in one half of the total number of
collisions participate two moving particles and the other half
involves one particle at rest.  Thus, in mean,
only one quarter of particles annihilated in collisions are at rest and the
three remaining quarters are formed by moving particles.
This simple heuristic argument suggests that the values
$p_0^\star=\frac{1}{4}$,
$p_+^\star=p_-^\star=\frac{3}{8}$ must play a special role:
if $p_0$ is less than $p_0^\star$, the system will asymptotically behave as in
the
two-velocity case (the
stationary particles will disappear before the annihilation of the moving
ones).
On the contrary, if $p_0$ is greater than $p_0^\star$, the moving particles
will
disappear first,
and asymptotically only particles at rest will be left.  In the limiting case
$p_0=p_0^\star$ the moving particles and the ones at rest disappear at the same
rate.  It will turn out that this intuitive argument is exact. We shall show it
using the approach
presented in~\cite{pias}, which permits to solve this three-velocity model
exactly.

This paper is organized as follows.  In section~\ref{sect2}, we introduce the
model and recall the main steps of the rigorous method~\cite{pias}.
An important quantity is
$S(v;t)$, the survival probability until the time $t$ of a particle moving with
velocity $v$.  In section~\ref{sect3}, we study the behavior of $S(v;t)$ and
compute the density and the time dependent velocity distribution in the
asymptotic regime $t\to\infty$. The value $p_0=\frac{1}{4}$ appears to be a
critical point, separating differents kinetic behaviors.  In addition, it is
shown that the dynamics is incompatible with a Boltzmann-like approximation
due to the appearance of strong velocity correlations between particles which
are nearest neighbors.
The case of a general discrete multi-velocity distribution is
examined in section~\ref{sect4}.  Finally, concluding remarks are given in
section~\ref{sect5}.

\section{The three-velocity model.}\label{sect2}

We assume that initially the particles are uniformly distributed in space,
according to the
Poisson law, without any correlations between their velocities.  Note that
other
distributions than Poisson could be considered as long as one is dealing with a
renewal process.  The process is thus translationally invariant and each
particle has initially the same probability density $\phi(v;0)$ (given by
equation~(\ref{eq:disv})) to move with velocity $v$.

The important characteristic of the annihilation dynamics is that only those
particles which suffered no collisions are present in the system.  They are
thus
found on their free trajectories.  Accordingly, the key quantity of the theory
is the survival probability $S(v;t)$ which has the product structure
\begin{equation}
S(v;t)=S^R(v;t)\,S^R(-v;t)\ ,
\end{equation}
where $S^R(v;t)$ is the probability for the absence of collision with the right
neighbor. As $S^R(-c;t)=1$, for  $t\geq 0$, we find
\begin{eqnarray}
S(+c;t)=S(-c;t)&=&S^R(+c;t)\ ,\\
S(0;t)&=&\left[S^R(0;t)\right]^2\ .
\end{eqnarray}
Initially, $S^R(+c;0)=S^R(0;0)=1$.

The density of particles with velocity $v$ at time $t$ is given by:
\begin{equation}
\sigma(v;t)=\sigma S(v;t)\phi(v;0)\ ,
\end{equation}
where $\sigma$ is the initial density.

Another important quantity in the approach~\cite{pias} is the distribution of
nearest neighbors.  Suppose that at time $t$ there is a particle moving with
velocity $v$. We denote by $\mu(x,u|0,v;t)$, the conditional probability
density
for finding its nearest neighbor to the right at a distance $x>0$, with
velocity
$u$ (for non-homogeneous systems the probability may depend upon the positions
of both particles, rather than on their relative distance only).  As we are
dealing here with symmetric initial velocity distribution, the following
relation holds for $\mu$:
\begin{equation}
\phi(v;0)\,S(v;t)\,\mu(x,u|0,v;t)=\phi(-u;0)\,S(-u;t)\,\mu(x,-v|0,-u;t)\ .
\label{eq:sim}
\end{equation}
The initial condition is given by
\begin{equation}
\mu(x,u|0,v;t=0)=\theta(x)\,\sigma\,{\rm e}^{-x\sigma}\phi(u;0)\ ,
\end{equation}
where $\theta(x)$ is the usual Heaviside unit step function.

We also define the density $\mu$ at contact by
\begin{equation}
\mu(0^+,u|0,v;t)=\lim_{\scriptstyle x\to 0\atop\scriptstyle x>0}\mu(x,u|0,v;t)
\ .
\end{equation}
This quantity plays a particular role, because it determines the
density of precollisional configurations.

The time evolution of $S^R(v;t)$ is given by:
\begin{eqnarray}
S^R(+c;t)&=&\exp\left\{-\int_0^t\!d\tau\left[c\,\mu(0^+,0|0,+c;\tau)+
2c\,\mu(0^+,-c|0,+c;\tau)\right]\right\}\label{eq:Splus}\\
S^R(0;t)&=&\exp\left\{-\int_0^t\!d\tau\,c\,\mu(0^+,-c|0,0;\tau)\right\}
\label{eq:Szero}
\end{eqnarray}
and the two-particle conditional probability density obeys to the following
closed equation~\cite{pias}
\[\left\{\frac{\partial}{\partial t}+z\,v_{21}+\frac{\dot{S}^R(v_1;t)}
{S^R(v_1;t)}-\frac{\dot{S}^R(v_2;t)}{S^R(v_2;t)}\right\}\tilde{\mu}(z,v_2|v_1;t)
-v_{21}\,\mu(0^+,v_{2}|0,v_1;t)=\]
\begin{equation}
=\int dv_3\!\int\!dv_4\,\tilde{\mu}(z,v_3|v_{1};t)\,\tilde{\mu}(z,v_2|v_4;t)
\,v_{34}\,\theta(v_{34})\,\mu(0^+,v_4|0,v_3;t)\ ,\label{eq:evol}
\end{equation}
where the dot stands for the time derivative, $v_{21}=v_2-v_1$, and we use
the definition
\begin{equation}
\tilde{\mu}(z,u|v;t)=\int_{0^+}^{\infty}\!dx\,{\rm e}^{-zx}\mu(x,u|0,v;t)\ .
\end{equation}
It should be stressed that this remarkable property of getting a rigorous
closed equation for the distribution $\mu$ comes from the main
characteristic of the annihilation process: only those particles which have not
interacted could
survive until time $t$. For more details see reference~\cite{pias}.
We can now start to solve equation~(\ref{eq:evol}) for our symmetric three
velocity case. Equation~(\ref{eq:evol}) can be rewritten in a matrix form:
\begin{equation}
\dot{\cal N}(z;t)+{\cal C}(z;t)={\cal N}(z;t)\,{\cal C}(z;t)\,{\cal N}(z;t)
\label{eq:mat}
\end{equation}
where $\cal N$ and $\cal C$ are two $3\times 3$ matrix:
\begin{equation}
{\cal N}(z;t)=\left(N_{ij}(z;t)\right)_{i,j=1,2,3}\,
\end{equation}
with
\begin{equation}\qquad\left.\begin{array}{rcl}
N_{11}(z;t)&=&\tilde{\mu}(z,+c|+c;t)\\ [.1cm]
N_{12}(z;t)&=&\tilde{\mu}(z,0|+c;t)\,{\rm e}^{-zct}\,S^R(+c;t)/S^R(0;t)\\
[.1cm]
N_{13}(z;t)&=&\tilde{\mu}(z,-c|+c;t)\,{\rm e}^{-2zct}\,S^R(+c;t)\\ [.1cm]
N_{21}(z;t)&=&\tilde{\mu}(z,+c|0;t)\,{\rm e}^{zct}\,S^R(0;t)/S^R(+c;t)\\ [.1cm]
N_{22}(z;t)&=&\tilde{\mu}(z,0|0;t)\\ [.1cm]
N_{23}(z;t)&=&\tilde{\mu}(z,-c|0;t)\,{\rm e}^{-zct}\,S^R(0;t)\\ [.1cm]
N_{31}(z;t)&=&\tilde{\mu}(z,+c|-c;t)\,{\rm e}^{2zct}/S^R(+c;t)\\ [.1cm]
N_{32}(z;t)&=&\tilde{\mu}(z,0|-c;t)\,{\rm e}^{zct}/S^R(0;t)\\ [.1cm]
N_{33}(z;t)&=&\tilde{\mu}(z,-c|-c;t)
\end{array}\qquad\right\}\label{eq:NM}
\end{equation}
and
\begin{equation}
{\cal C}(z;t)=\left(\begin{array}{ccc}
0&C_{12}(z;t)&C_{13}(z;t)\\0&0&C_{23}(z;t)\\0&0&0
\end{array}\right)\ ,\label{eq:strc}
\end{equation}
with
\begin{equation}\qquad\left.\begin{array}{rcl}
C_{12}(z;t)&=&c\,\mu(0^+,0|0,+c;t)\,{\rm e}^{-zct}\,S^R(+c;t)/S^R(0;t)\\ [.1cm]
C_{13}(z;t)&=&2c\,\mu(0^+,-c|0,+c;t)\,{\rm e}^{-2zct}\,S^R(+c;t)\\ [.1cm]
C_{23}(z;t)&=&c\,\mu(0^+,-c|0,0;t)\,{\rm e}^{-zct}\,S^R(0;t)\ .
\end{array}\qquad\right\}\label{eq:Cij}
\end{equation}

The initial values of ${\cal N}$ and ${\cal C}$ are:
\begin{eqnarray}
{\cal N}(z;0)&=&\frac{\sigma}{z+\sigma}\left(
\begin{array}{ccc}p_+&p_0&p_+\\p_+&p_0&p_+\\p_+&p_0&p_+\end{array}
\right)\ ,\\
{\cal C}(z;0)&=&\left(
\begin{array}{ccc}0&c\sigma p_0&2c\sigma p_+\\0&0&c\sigma p_+\\0&0&0\end{array}
\right)\ .
\end{eqnarray}
In spite of the fact that equation~(\ref{eq:mat}) looks very simple, it should
be noticed that it is a matrix differential equation for which
the general solution is not known. However, the particular structure of
 $\cal C$ makes
it solvable. The method of solution is described in appendix~\ref{appA}. We
find
\begin{equation}\qquad\left.\begin{array}{rcl}
N_{11}(z;t)&=&{\displaystyle \frac{p_+\sigma}{D(z;t)}\left(1-\frac{p_0}{2p_+}
\left[\int_0^t\!d\tau\,C_{23}(z;\tau)\right]^2\right)}\\ [.4cm]
N_{12}(z;t)&=&{\displaystyle \frac{p_0\sigma}{D(z;t)}\left(1-\frac{p_0}{2p_+}
\left[\int_0^t\!d\tau\,C_{23}(z;\tau)\right]^2\right)
-\frac{p_0}{p_+}\int_0^t\!d\tau\,C_{23}(z;\tau)}\\[.4cm]
N_{13}(z;t)&=&{\displaystyle \frac{p_+\sigma}{D(z;t)}\left(1-\frac{p_0}{2p_+}
\left[\int_0^t\!d\tau\,C_{23}(z;\tau)\right]^2\right)^2
-\int_0^t\!d\tau\,C_{13}(z;\tau)}\\[.4cm]
N_{21}(z;t)&=&{\displaystyle \frac{p_+\sigma}{D(z;t)}}\\[.4cm]
N_{22}(z;t)&=&{\displaystyle \frac{p_0\sigma}{D(z;t)}}\\[.4cm]
N_{23}(z;t)&=&{\displaystyle \frac{p_+\sigma}{D(z;t)}\left(1-\frac{p_0}{2p_+}
\left[\int_0^t\!d\tau\,C_{23}(z;\tau)\right]^2\right)
-\int_0^t\!d\tau\,C_{23}(z;\tau)}\\[.4cm]
N_{31}(z;t)&=&{\displaystyle \frac{p_+\sigma}{D(z;t)}}\\[.4cm]
N_{32}(z;t)&=&{\displaystyle \frac{p_0\sigma}{D(z;t)}}\\[.4cm]
N_{33}(z;t)&=&{\displaystyle \frac{p_+\sigma}{D(z;t)}\left(1-\frac{p_0}{2p_+}
\left[\int_0^t\!d\tau\,C_{2,3}(z;\tau)\right]^2\right)}
\end{array}\qquad\right\}\label{eq:solmat}
\end{equation}
where
\begin{equation}
D(z;t)={z+\sigma-2p_0\sigma\int_0^t\!d\tau\,C_{23}(z;\tau)-p_+\sigma\int_0^t
\!d\tau\,C_{13}(z;\tau)}\ .\label{eq:Dzt}
\end{equation}

The Laplace transformed conditional probabilities $\tilde{\mu}$ are obtained
from equations~(\ref{eq:NM}). Thus we found the exact solution to the kinetic
equation~(\ref{eq:evol}). However, this is as yet an implicit solution, that is
to say, the values of $\tilde{\mu}$ are expressed in terms of the density at
contact $\mu(0^+,u|0,v;t)$ ($u<v$).  To obtain some information about physical
quantities, we have to solve the consistency equations which express the three
non-vanishing densities at contact in terms of function $\tilde{\mu}$:
\begin{equation}
\mu(0^+,u|0,v;t)=\int\limits_{\gamma-i\infty}^{\gamma+i\infty}\frac{dz}
{2i\pi}\tilde{\mu}(z,u|v;t)\qquad (u<v). \label{eq:cons}
\end{equation}

Up to this point, the genralization to a non-symmetric three-velocity
distribution is straightforward.

\section{Long time behavior of the system.}\label{sect3}

Instead of writing the consistency equation for the density at contact, we
shall
write them for the survival probabilities.  Remembering that only two densities
at contact are independent (see equation~(\ref{eq:sim})) and that (from
equations~(\ref{eq:Splus}) and~(\ref{eq:Szero}))
\begin{eqnarray}
\dot{S}^R(0;t)&=&-c\mu(0,-c|0,0;t)\,S^R(0;t)\\
\dot{S}^R(+c;t)&=&\left[-c\mu(0,0|0,+c;t)-2c\mu(0,-c|0,+c;t)\right]S^R(+c;t)\ ,
\end{eqnarray}
we obtain, using equation~(\ref{eq:cons}), the following two equations:
\begin{eqnarray}
\dot{S}^R(0;t)&=&-cp_+\sigma\int\limits_{\gamma-i\infty}^{\gamma+i\infty}
\frac{dz}{2i\pi}{\rm e}^{zct}\frac{1-\frac{p_0}{2p_+}\left[\int_0^t\!d\tau\,
C_{23}(z;\tau)\right]^2}{D(z;t)}\label{eq:cons1t}\\
\dot{S}^R(+c;t)&=&\frac{p_0}{p_+}S^R(0;t)\dot{S}^R(0;t)\label{eq:cons2t}\\
&&-2cp_+\sigma\int\limits_{\gamma-i\infty}^{\gamma+i\infty}\frac{dz}{2i\pi}
{\rm e}^{2zct}\frac{\left(1-\frac{p_0}{2p_+}\left[\int_0^t\!d\tau\,
C_{23}(z;\tau)\right]^2\right)^2}{D(z;t)}\nonumber
\end{eqnarray}
Solving this two equations seems to be extremely difficult, principally because
of the time dependence in $D(z;t)$ and in the integrals over $\tau$. However it
can be shown (see appendix~\ref{appB}) that the only relevant time dependence
occurs in the exponential factor. By this we mean that
equations~(\ref{eq:cons1t}) and~(\ref{eq:cons2t}) can be exactly replaced by
the
following two relations:
\begin{eqnarray}
\dot{S}^R(0;t)&=&-cp_+\sigma\int\limits_{\gamma-i\infty}^{\gamma+i\infty}
\frac{dz}{2i\pi}{\rm e}^{zct}\frac{1-\frac{p_0}{2p_+}\left[\int_0^{\infty}
\!d\tau\,C_{23}(z;\tau)\right]^2}{D(z;\infty)}\label{eq:cons1inf}\\
\dot{S}^R(+c;t)&=&\frac{p_0}{p_+}S^R(0;t)\dot{S}^R(0;t)\label{eq:cons2inf}\\
&&-2cp_+\sigma\int\limits_{\gamma-i\infty}^{\gamma+i\infty}\frac{dz}{2i\pi}
{\rm e}^{2zct}\frac{\left(1-\frac{p_0}{2p_+}\left[\int_0^{\infty}\!d\tau
\,C_{23}(z;\tau)\right]^2\right)^2}{D(z;\infty)}\nonumber
\end{eqnarray}
This remarkable fact allows us to use again the Laplace transform to suppress
the inverse Laplace integral and arrive at much simpler relations:
\begin{eqnarray}
T(z)&\equiv&\int_0^{\infty}\!dt\,{\rm e}^{-zct}\dot{S}^R(0;t)\nonumber\\
&=&-p_+\sigma\frac{1-\frac{p_0}{2p_+}T^2(z)}{z+\sigma+2p_0\sigma T(z)+p_+
\sigma U(z)}\\
U(z)&\equiv&\int_0^{\infty}\!dt\,{\rm e}^{-2zct}\left[\dot{S}^R(+c;t)
-\frac{p_0}{p_+}\dot{S}^R(0;t)S^R(0;t)\right]\nonumber\\
&=&-p_+\sigma\frac{\left[1-\frac{p_0}{2p_+}T^2(z)\right]^2}
{z+\sigma+2p_0\sigma T(z)+p_+\sigma U(z)}\ ,
\end{eqnarray}
where equations~(\ref{eq:Cij}) and~(\ref{eq:Dzt}) have been used.
{}From this system of equations it follows that $T(z)$ satisfies the
quartic equation
\begin{equation}
p_0\,T^4(z)-\left(2p_0+1\right)T^2(z)-2\left(\frac{z}{\sigma}+1\right)T(z)
+p_0-1=0\label{eq:quartic}\ ,
\end{equation}
$U(z)$ being given by
\begin{equation}
U(z)=T(z)-\frac{p_0}{2p_+}T^3(z)\ .\label{eq:SfromT}
\end{equation}
It is hopeless to try to Laplace invert the solution of
equation~(\ref{eq:quartic}) to find $\dot{S}^R(0;t)$. However, the asymptotic
regime of the survival probability can be determined by considering $T(z)$ for
$z$ in the neighborhood of 0. For $z=0$, one knows that $T(0)$ is given by
\begin{equation}
T(0)=\int_0^{\infty}\!dt\,\dot{S}^R(0;t)=S^R(0,\infty)-1
\end{equation}
and that $S^R(0,\infty)\in[0,1]$. This condition allows us to identify the
physically acceptable solution of the above quartic
equation~(\ref{eq:quartic}).
We have to distinguish between two different situations:
\begin{itemize}
\item $p_0\leq\frac{1}{4}$

The only acceptable solution is $T(0)=-1$, corresponding to an asymptotically
empty stationary
state $S^R(+c;\infty)=S^R(0;\infty)=0$.

\item $p_0\geq\frac{1}{4}$

We have two possible solutions:  $T(0)=-1$ and $T(0)=1-\frac{1}{\sqrt{p_0}}$.
However, $S^R(0;t)$ is a  continuous decreasing function of time with initial
value 1. Thus the stationary value $2-\frac{1}{\sqrt{p_0}}$ will be reached
first. Accordingly, $T(0)=1-\frac{1}{\sqrt{p_0}}$ is the relevant value.
\end{itemize}

We can now study the behavior of $T(z)$ near zero, by introducing the quantity
$\epsilon(z)\equiv T(z)-T(0)$ in the quartic equation~(\ref{eq:quartic}). Five
cases have to be distinguished:
\begin{itemize}
\item $p_0=0$

This is the bimodal velocity distribution already investigated~\cite{pias,EF}.
The quartic equation~(\ref{eq:quartic}) simplifies to the following quadratic
equation for $U(z)$:
\begin{equation}
U^2(z)+2(\frac{z}{\sigma}+1)U(z)+1=0\ .
\end{equation}
By solving and Laplace inverting it, one recovers the Elskens and Frisch
results~\cite{EF}.

\item $0<p_0<\frac{1}{4}$

Now $\epsilon(z)=T(z)+1$ obeys the quartic equation:
\begin{equation}
p_0\epsilon^4(z)-4p_0\epsilon^3(z)+(4p_0-1)\epsilon^2(z)-
2\frac{z}{\sigma}(\epsilon(z)-1)=0\ .\label{eq:quaeps}
\end{equation}
As we are interested in the limit $z\to 0$, and as $\lim_{z\to 0}
\epsilon(z)=0$, we can neglect the terms of order $\epsilon^4$ and
$\epsilon^3$ with respect to $\epsilon^2$, and $\epsilon$ with respect to $1$.
By Laplace inverting and integrating, we obtain the following asymptotic
behavior ($t\to\infty$):
\begin{eqnarray}
S^R(0;t)&=&\sqrt{\frac{2}{(1-4p_0)\pi c\sigma t}}\left\{1+
{\cal O}\left([1-4p_0]^{-3}[c\sigma t]^{-1}\right)\right\}\ ,\\
S^R(+c;t)&=&\frac{1}{1-p_0}\sqrt{\frac{1-4p_0}{\pi c\sigma t}}\left\{1+
{\cal O}\left([1-4p_0]^{-3/2}[c\sigma t]^{-1/2}\right)\right\}\ .
\end{eqnarray}
For $t\to\infty$, the densities are then given by:
\begin{eqnarray}
\sigma(0;t)&=&\frac{2p_0}{(1-4p_0)c\pi t}\left\{1+
{\cal O}\left([1-4p_0]^{-3}[c\sigma t]^{-1}\right)\right\}\ ,\\
\sigma(+c;t)&=&\sqrt{\left(\frac{1}{4}-p_0\right)\frac{\sigma}{c\pi
t}}\left\{1+
{\cal O}\left([1-4p_0]^{-3/2}[c\sigma t]^{-1/2}\right)\right\}\ .
\end{eqnarray}
The corrections to scaling have been obtained by a careful study of the
solution
of the quartic equation~(\ref{eq:quartic}).

The prefactor $(4p_0-1)$ in the $\epsilon^2$ term in
equation~(\ref{eq:quaeps}) indicates that for $p_0$ near to $\frac{1}{4}$  we
should expect important crossover effect.  This is confirmed by the amplitude
of
the correction to scaling for $\sigma(0;t)$ which contains a term proportional
to $(1-4p_0)^{-4}$. As a consequence, it may be very difficult to extract the
true asymptotic behavior from experimental or numerical data.  However, we see
that for very long times the system is driven towards the bimodal case.

\item $p_0=\frac{1}{4}$

Here again, $\epsilon(z)=T(z)+1$.  The coefficient of the $\epsilon^2$ term
vanishes and the term of order $\epsilon^3$ cannot be neglected.  We are thus
left with the cubic equation:
\begin{equation}
\epsilon^3(z)-2\frac{z}{\sigma}=0,\qquad {\rm when}\ z\to 0
\end{equation}
whose physically relevant solution leads to the asymptotic behavior (for
$t\to\infty$):
\begin{eqnarray}
S^R(0;t)&=&\frac{1}{\Gamma(2/3)}\left(\frac{2}{c\sigma t}\right)^{1/3}
\left\{1+{\cal O}\left([c\sigma t]^{-1/3}\right)\right\}\ ,\\
S^R(+c;t)&=&\left[\frac{1}{3}\left(\frac{2^{1/3}}
{\Gamma(2/3)}\right)^2+\frac{1}{\Gamma(1/3)}\right]\left(\frac{1}{c\sigma t}
\right)^{2/3}\left\{1+{\cal O}\left([c\sigma t]^{-1/3}\right)\right\}\ .
\end{eqnarray}
For the density one gets ($t\to\infty$):
\begin{eqnarray}
\sigma(0;t)&=&\frac{\sigma}{4\Gamma^2(2/3)}\left(\frac{2}{c\sigma t}
\right)^{2/3}\left\{1+{\cal O}\left([c\sigma t]^{-1/3}\right)\right\}\ ,\\
\sigma(+c;t)&=&\frac{3\sigma}{8}\left[\frac{1}{3}\left(\frac{2^{1/3}}
{\Gamma(2/3)}\right)^2+\frac{1}{\Gamma(1/3)}\right]\left(\frac{1}{c\sigma t}
\right)^{2/3}\left\{1+{\cal O}\left([c\sigma t]^{-1/3}\right)\right\}\ .
\end{eqnarray}
In addition,
\begin{equation}
\lim_{t\to\infty}\frac{\sigma(+c;t)}{\sigma(0;t)}=\frac{1}{2}+\frac{3}{2}
\frac{\Gamma^2(2/3)}{2^{2/3}\Gamma(1/3)}\simeq 1.15\ .
\end{equation}

\item $\frac{1}{4}<p_0<1$

Now $\epsilon(z)=T(z)-1+\frac{1}{\sqrt{p_0}}$. The terms of order
$\epsilon^4$ and $\epsilon^3$ can be neglected with respect to the term
$\epsilon^2$ and variable $z$  with respect to 1. Hence:
\begin{eqnarray}
\epsilon(z)&=&\frac{2\sqrt{p_0}-1}{p_0(5-2\sqrt{p_0})}\\
&-&\sqrt{\frac{2(1-\sqrt{p_0})}{p_0(2\sqrt{p_0}-1)(5-2\sqrt{p_0})}}
\left[\frac{(2\sqrt{p_0}-1)^3}{2(1-\sqrt{p_0})(5-2\sqrt{p_0})}+
\frac{z}{\sigma}\right]^{1/2}\ ,\nonumber
\end{eqnarray}
which leads to an exponential asymptotic behavior for $S^R(0;t)-2+\frac{1}
{\sqrt{p_0}}$ and also for $S^R(+c;t)$. Thus, for $t\to\infty$
\begin{eqnarray}
\sigma(0;t)&\simeq&\sigma\left(2\sqrt{p_0}-1\right)^2+2\sigma A
\frac{{\rm e}^{-c\sigma ut}}{(c\sigma t)^{3/2}}\ ,\\
\sigma(+c;t)&\simeq&\sigma A\frac{{\rm e}^{-c\sigma ut}}{(c\sigma t)^{3/2}}\ ,
\end{eqnarray}
where
\begin{eqnarray*}
A&=&\sqrt{\frac{2p_0(1-\sqrt{p_0})(5-2\sqrt{p_0})}{\pi(2\sqrt{p_0}-1)}}
\frac{1-\sqrt{p_0}}{(2\sqrt{p_0}-1)^2}\ ,\\
u&=&\frac{(2\sqrt{p_0}-1)^3}{2p_0(1-\sqrt{p_0})(5-2\sqrt{p_0})}\ .
\end{eqnarray*}
In addition,
\begin{equation}
\lim_{t\to\infty}\frac{\sigma(0;t)-\sigma(0;\infty)}{\sigma(+c;t)}=2\ ,
\end{equation}
indicating that in the long time regime the collisions between pairs of moving
particles disappear, and only collisions involving a particle at rest and a
moving one can be observed.
\item $p_0=1$

In this trivial limiting case, in which there is no collisions,
the physical solution is  $T(z)\equiv 0$, and the density remains constant.
\end{itemize}

As anticipated on heuristic arguments, we have shown that $p_0=\frac{1}{4}$ is
playing a special role separating the different kinetic regimes.

Another important quantity that can be computed exactly concerns the
correlations between the velocities of the colliding particles.  Although at
time $t=0$ these velocities are uncorrelated, annihilation dynamics creates
strong correlations between them during the time evolution.  This has the
important consequence to exclude a Boltzmann-like approximation. It is clearly
seen on ${\bar {w}}(v;t)$, the mean velocity of the right nearest neighbor of a
particle moving with velocity $v$:
\begin{equation}
{\bar {w}}(v;t)=c\tilde{\mu}(0,+c|v;t)-c\tilde{\mu}(0,-c|v;t)\ .
\end{equation}
Using equations~(\ref{eq:solmat}) we obtain for example:
\[
{\bar {w}}(-c;t)=\frac{c}{D(0;t)}\left[p_+S^R(+c;t)-p_++\frac{p_0}{2}
\left(\int_0^t\!d\tau\,C_{23}(0;\tau)\right)^2\right]\ ,
\]
with $D(0;t)$ given by equation~(\ref{eq:Dzt}) and $C_{23}(0;t)$ by
equation~(\ref{eq:Cij}).
Our results show that pairs of nearest neighbors have the tendency to
align their velocities and propagate in the same direction.  Indeed, in the
limit $t\to\infty$, one finds for $p_0<\frac{1}{4}$:
\begin{equation}
\bar{w}(v;\infty)=\left\{\begin{array}{ll}\pm c&\ \mbox{if $v=\pm c$}\\
{\displaystyle c\,(\sqrt{2}-1)\frac{1-4p_0}{1-2p_0}}&\ \mbox{if $v=0$,}
\end{array}\right .
\end{equation}
for $p_0=\frac{1}{4}$
\begin{equation}
{\bar {w}}(v;\infty)=v
\end{equation}
and for $p_0>\frac{1}{4}$
\begin{equation}
{\bar {w}}(v;\infty)=v (p_0^{-1/2}-1)\ .
\end{equation}

\section{General discrete velocity distribution}\label{sect4}

In this section, we want to consider the case of a general discrete velocity
distribution. Suppose that the initial velocity distribution is given by
\begin{equation}
\phi(v;0)=\sum_{k=1}^n p_k\,\delta(v-v_k)\ ,
\end{equation}
with
\[\sum_{k=1}^np_k=1\ .\]
As for the three-velocity case, we assume initially random spatial distribution
and no correlations between the velocities.  We can generalize some of our
formulae in a straightforward way:  the probability for the
absence of collisions with the right neighbor is given by:
\begin{equation}
S^R(v;t)=\exp\left\{\int_0^t\!d\tau\sum_{k=1}^n (v-v_k)\,\theta(v-v_k)
\mu(0^+,v_k|0,v;\tau)\right\}\ .
\end{equation}
If we define ($i,j=1,\ldots,n$):
\begin{eqnarray}
N_{ij}(z;t)&=&\tilde{\mu}(z,v_j|v_i;t)\exp\left\{-zv_{ij}t+\int_0^t\!d\tau
\sum_{k=1}^3\left[B_{ik}(\tau)-B_{jk}(\tau)\right]\right\}\\
C_{ij}(z;t)&=&B_{ij}(t)\exp\left\{-zv_{ij}t+\int_0^t\!d\tau\sum_{k=1}^3
\left[B_{ik}(\tau)-B_{jk}(\tau)\right]\right\}\,
\end{eqnarray}
where
\[
B_{ij}(t)=v_{ij}\,\theta(v_{ij})\,\mu(0,v_j|0,v_i;t)\ ,
\]
the evolution equation for the Laplace transformed conditional probability
$\tilde{\mu}$ (equation~\ref{eq:evol}) can be put in the same form as
equation~(\ref{eq:mat}), but with a $n\times n$ matrix.  Once again, the
particular structure of the matrix $\cal C$ will help us in solving this
equation:  assuming that ${\cal N}(z;t)-\cal I$ is invertible (this can be
explicitly checked for $t=0$ and verified a posteriori for later times), where
$\cal I$ is the $n\times n$ unit matrix, we define
\begin{equation}
{\cal P}(z;t)=\left({\cal N}(z;t)-{\cal I}\right)^{-1}+\frac{{\cal I}}{2}\ .
\label{eq:change}
\end{equation}
Then, equation~(\ref{eq:mat}) reads
\begin{equation}
\dot{\cal P}(z;t)=-{\cal P}(z;t){\cal C}(z;t)-{\cal C}(z;t){\cal P}(z;t)\ .
\label{eq:mat2}
\end{equation}
One way to solve this equation is to remember that $C_{ij}(z;t)=0$ for $i\geq
j$
($i,j=1,\ldots,n$). Hence, equation~(\ref{eq:mat2}) has the following
structure:
\begin{equation}
-\dot{P}_{ij}(z;t)=\sum_{k=1}^{j-1}P_{ik}\,C_{kj}+\sum_{k=i+1}^nC_{ik}\,P_{kj}
\end{equation}
and thus when $i=n,j=1$ it takes the form
\[\dot{\cal P}_{n,1}(z;t)=0\ .\]
In addition, the $(n-1,1)$ and $(n,2)$ equations are
\begin{eqnarray*}
-\dot{P}_{n-1,1}(z;t)&=&C_{n-1,n}(z;t)P_{n,1}(z;t)\\
-\dot{P}_{n,2}(z;t)&=&C_{1,2}(z;t)P_{n,1}(z;t)\ ,
\end{eqnarray*}
and can be solved once $P_{n,1}(z;t)$ is known.  One can easily see that the
equations for $(n-2,1)$, $(n-1,2)$ and $(n,3)$ are expressed in term of
$P_{n,1}$, $P_{n-1,1}$ and $P_{n,2}$, allowing us to complete the solution.
The
process is iterated until solving equation ($1,n$) and determining entirely the
matrix ${\cal P}(z;t)$.  The $n^2$ Laplace transformed conditional
probabilities are obtained straightforward by inverting the
formula~(\ref{eq:change}).  Again we have an implicit solution and we have to
solve the $n(n-1)/2$ consistency equations (see equation~(\ref{eq:cons})) to
obtain physical information on the system.  Although detailed calculations can
be very tedious, there should not be any conceptual difficulties.

\section{Concluding remarks}\label{sect5}

We have shown in this paper that new exact predictions on ballistic
annihilation
kinetics can be obtained in the framework of the new approach~\cite{pias}. It
is
remarkable that, for discrete velocity distributions, the non-linear
integro-differential equation governing the two particle conditional
probability
can be solved exactly.

For $p_0\leq\frac{1}{4}$, not only the leading power laws are obtained in the
asymptotic regime, $t\to\infty$, but also the amplitudes and the corrections to
scaling. In particular, we have shown that important corrections to scaling
occur when $p_0$ is in the neighborhood of $\frac{1}{4}$. For
$p_0>\frac{1}{4}$, exponential behavior is obtained for long times. The results
of our numerical simulations and those of Redner et al.~\cite{BRL,R} are well
explained by our exact theory. In addition, it is clearly shown that
Boltzmann-like approximations fail, because the annihilation dynamics favorizes
configurations in which the nearest neighbors have the same velocity.

Several extensions of this work are possible. In particular, the case of
continuous velocity distributions for which qualitatively different behavior
may be expected is under investigation.

\appendix{\section{On the matrix equation}\label{appA}

The solution of the matrix equation~(\ref{eq:mat}) is presented here.  To solve
it, we can write ${\cal N}$ as
\begin{equation}
{\cal N}(z;t)=\left(
\begin{array}{cc}{\cal N}_1(z;t)&{\cal N}_2(z;t)\\{\cal N}_3(z;t)&{\cal N}_4
(z;t)\end{array}\right)
\end{equation}
where ${\cal N}_1$ is a $1\times 2$ matrix, ${\cal N}_2$ a $1\times 1$,
${\cal N}_3$ a $2\times 2$ and ${\cal N}_4$ a $2\times 1$. In this
decomposition, ${\cal C}$ is given by
\begin{equation}
{\cal C}(z;t)=\left(
\begin{array}{cc}{\cal E}_1(z;t)&{\cal E}_2(z;t)\\0&{\cal E}_3(z;t)
\end{array}\right)\end{equation}
where
\begin{eqnarray*}
{\cal E}_1(z;t)&=&\left(\begin{array}{cc}0&C_{12}(z;t)\end{array}\right)\ ,\\
{\cal E}_2(z;t)&=&\left(\begin{array}{c}C_{13}(z;t)\end{array}\right)\ ,\\
{\cal E}_3(z;t)&=&\left(\begin{array}{c}C_{23}(z;t)\\0\end{array}\right)\ .
\end{eqnarray*}
However, to perform the matrix product ${\cal N}\,{\cal C}\,{\cal N}$, we
have to write ${\cal C}$ in a different way, namely:
\begin{equation}
{\cal C}(z;t)=\left(
\begin{array}{cc}0&{\cal E}(z;t)\\0&0\end{array}
\right)\ ,\end{equation}
where ${\cal E}$ is a $2\times 2$ matrix
\[{\cal E}(z;t)=\left(\begin{array}{cc}
C_{1,2}(z;t)&C_{1,3}(z;t)\\0&C_{2,3}(z;t)\end{array}\right)\ .\]

With this decomposition, the matrix equation~(\ref{eq:mat}) becomes
\begin{eqnarray}
\left(\begin{array}{cc}
\dot{\cal N}_1(z;t)&\dot{\cal N}_2(z;t)\\ \dot{\cal N}_3(z;t)&\dot{\cal N}_4
(z;t)\end{array}\right)&
+&\left(\begin{array}{cc}{\cal E}_1(z;t)&{\cal E}_2(z;t)\\
0&{\cal E}_3(z;t)\end{array}\right)=\\
&=&\left(\begin{array}{cc}{\cal N}_1(z;t)
{\cal E}(z;t){\cal N}_3(z;t)&{\cal N}_1(z;t){\cal E}(z;t){\cal N}_4(z;t)\\
{\cal N}_3(z;t){\cal E}(z;t){\cal N}_3(z;t)&{\cal N}_3(z;t){\cal E}(z;t)
{\cal N}_4(z;t)\end{array}\right)\ .\nonumber
\end{eqnarray}
Although ${\cal N}_3$ is not invertible, we can find a solution for the
equation
\begin{equation}
\dot{\cal N}_3(z;t)={\cal N}_3(z;t){\cal E}(z;t){\cal N}_3(z;t)
\end{equation}
using the following Ansatz:
\begin{equation}
{\cal N}_3(z;t)={\cal A}(z)\left[{\cal I}-{\cal B}(z;t){\cal A}(z)\right]^{-1}
\end{equation}
where ${\cal I}$ is the unit $2\times 2$ matrix. Once we have obtained
${\cal N}_3$, we can readily solve the three other matrix equations and
thus find ${\cal N}$.

\section{On the consistency equations}\label{appB}

In this appendix we show how to justify equation~(\ref{eq:cons1inf}) and
(\ref{eq:cons2inf}) starting from equation~(\ref{eq:cons1t})
and~(\ref{eq:cons2t}).

A way of demonstration could be to simply subtract
equations~(\ref{eq:cons1inf})
and~(\ref{eq:cons1t}) and to verify by explicit integration in the complex
plane, that the result is zero.  Instead of going through all this tedious
algebra, we prefer to give a simple physical argument.

We define $\tilde{A}_1(z;t)$ by
\begin{equation}
\tilde{\mu}(z,+c|+c;t)=p_+\sigma\tilde{A}_1(z;t)\ .\label{eq:A1t}
\end{equation}
{}From the kinetic equation~(\ref{eq:evol}), we have
\begin{equation}
\dot{\tilde{\mu}}(z,+c|+c;t)=\sum_{i=1}^3\sum_{j=1}^3
\tilde{\mu}(z,v_i|+c;t)\tilde{\mu}(z,+c|v_j;t)
v_{ij}\theta(v_{ij})\mu(0^+,v_j|0,v_i;t)\ .\label{eq:evcc}
\end{equation}
Now, the right hand side of equation~(\ref{eq:evcc}) represents the variation
of
$\tilde{\mu}$ due to the mutual annihilation of particles originally separating
the pair ($+c$, $+c$) (see reference~\cite{pias} for the interpretation of
different terms in equation~(\ref{eq:evol})).  It follows that the inverse
Laplace transform
\begin{equation}
A_1(x;t)=\int\limits_{\gamma-i\infty}^{\gamma+i\infty}\frac{dz}{2i\pi}\,
{\rm e}^{xz}\tilde{A}_1(z;t)
\end{equation}
represents the probability that the particles present initially in an interval
of length $x$, separating two particles with velocity $+c$, disappear through
ballistic annihilation before time $t$ (the prefactor $p_+\sigma$ in
equation~(\ref{eq:A1t}) is the initial density of the right neighbors moving
with velocity $+c$).  As the colliding pairs move at least with a relative
velocity $c$, all sequences of encounters contributing to $A_1(x;t)$ are
accomplished at the moment $t^{\star}=x/c$.  Therefore $A_1(x;t)$ does not
depend on time (for fixed $x$) in the region $t>t^{\star}$.

Now we remember that
\begin{equation}
\tilde{A}_1(z;t)=\frac{1-\frac{p_0}{2p_+}
\left[\int_0^t\!d\tau\,C_{23}(z;\tau)\right]^2}{D(z;t)}\ ,
\end{equation}
hence the equation~(\ref{eq:cons1t}) reads:
\begin{equation}
\dot{S}^R(0;t)=-cp_+\sigma\int\limits_{\gamma-i\infty}^{\gamma+i\infty}
\frac{dz}{2i\pi}{\rm e}^{zct}\tilde{A}_1(z;t)\ .\label{eq:SA1}
\end{equation}
In the term on the right hand side of equation~(\ref{eq:SA1}), the inverse
Laplace transformation yields the value of function $A_1(x;t)$ at the point
$x=ct$.  So we can replace $\tilde{A}_1(z;t)$ in equation~(\ref{eq:SA1}) by
$\tilde{A}_1(z;\infty)$, without changing the value of the integral.  Hence we
obtain, equation~(\ref{eq:cons1inf}).

The equation~(\ref{eq:cons2t}) can be handled in the same manner: we first
define
\begin{equation}
p_+\sigma\tilde{A}_2(z;t)=\tilde{\mu}(z,-c|+c;t)\,{\rm e}^{-2zct}\,S^R(+c;t)+
\int_0^t\!dt\,C_{13}(z;t)\ .
\end{equation}
Then, by inspecting the evolution equation for $\tilde{\mu}(z,-c|+c;t)$ we
remark that $A_2(x;t)$, the inverse Laplace transform of $\tilde{A}_2(z;t)$,
represents the probability that all particles present initially in an interval
of length $x$, separating one particle of velocity $+c$ on the left and one of
velocity $-c$ on the right, disappear through ballistic annihilation before
time
$t$.  It follows that $A_2(x;t)$ does not depend on the time (for fixed x) in
the region $t>t^{\star}/2$ (as the relative velocity between the colliding
pairs must be $2c$).  Finally, the expression
\begin{equation}
\tilde{A}_2(z;t)=\frac{\left(1-\frac{p_0}{2p_+}
\left[\int_0^t\!d\tau\,C_{23}(z;\tau)\right]^2\right)^2}{D(z;t)}
\end{equation}
leads to the expecting conclusion.
}


\begin{references}
\bibitem[1]{ack_uno} Supported by the Swiss National Foundation.
\bibitem[2]{ack_due} Work partly done in Geneva. Supported by the Swiss
National
Foundation.
\bibitem[3]{ack_tre} The hospitality at the Department of Theoretical Physics
of the University of Geneva is greatly acknowledged.

\bibitem{EF} Y. Elskens and H.L. Frisch, Phys. Rev. {\bf A31}, 3812 (1985)
\bibitem{KS} J. Krug and H. Spohn, Rev. {\bf A38}, 4271 (1988)
\bibitem{BRL} E. Ben-Naim, S. Redner and F. Leyvraz, Phys. Rev. Lett.
{\bf 70}, 1890 (1993)
\bibitem{R} S. Redner, to appear in {\it Proceedings of the 2nd International
Colloquium on Quantum Field Theory and Stochastic Processes} (World
Scientific).
\bibitem{pias} J. Piasecki,  submitted to Phys. Rev. {\bf E} (1994)
\end{references}
\end{document}